\documentclass[a4paper]{article}
\RequirePackage[english]{babel}
\RequirePackage[latin1]{inputenc}
\RequirePackage[T1]{fontenc}
\RequirePackage{mathrsfs}
\RequirePackage{amsmath}
\RequirePackage{amssymb}
\RequirePackage{amsbsy}
\RequirePackage{bm}
\begin{document}
\title{\bf{On the consistency of Constraints in\\ Matter Field Theories}}
\author{Luca Fabbri\\ 
\footnotesize D\'{e}partement de Physique, Universit\'{e} de Montr\'{e}al, CANADA\\
\footnotesize INFN \& Dipartimento di Fisica, Universit\`{a} di Bologna, ITALY}
\date{}
\maketitle
\begin{abstract}
We consider how the principles of causality and equivalence restrict the background in which matter field theories are defined; those constraints develop in restrictions for these matter field theories: the simplest matter field theory aside, all other less simple matter field theories are too complex therefore resulting to be inconsistent in general instances.
\end{abstract}
\section*{Introduction}
The Principles of Causality and Equivalence are the two fundamental principles that define the properties of the geometry, because they imply the light-cone structure be preserved in the unique system of reference that is locally in free-fall; then metric connections with completely antisymmetric Cartan torsion tensors have to be required for consistency \cite{h,a-l,m-l,f/1,f/2}. 

However if matter fields have higher-spin then the problem of being boosted out of the light-cone during free-fall or lacking free-fall at all arises, and therefore such matter fields with higher-spin will be characterized by inconsistencies; on the other hand the definition of higher-spin matter fields can avoid inconsistencies if constraints are imposed: this procedure is a commonly used prescription discussed in what is known as the Velo-Zwanziger problem \cite{v-z/1,v-z/2,p-r}.

However the Velo-Zwanziger problem in its original form considers connections that are non-trivial only for the presence of gauge fields, and in following papers the Velo-Zwanziger problem generalized up to connections that also have the metric tensor beside the torsion tensor is discussed \cite{d-z,d-w,u-v,n}; beyond the analysis of this generalized Velo-Zwanziger problem, in this paper we will also consider the torsion-spin coupling imposing the complete antisymmetry of the spin tensor itself \cite{F}: the constraints due to the Velo-Zwanziger analysis will not only be strengthened by the presence of the torsional back-reaction of the torsion-spin coupling of the fields with themselves but furthermore they will also be accompanied by additional constraints on the complete antisymmetry of their spin. We will see that these constraints for the matter fields will eventually result in severe restrictions that cannot be met by too complex matter field theories, and therefore when a matter field theory has a too complex structure then it will turn out to be inconsistent in the most general situation possible.
\section{Constraints in Matter Field Theories}
In a given geometry, the metric structure is given in terms of two symmetric metric tensors $g_{\alpha\beta}$ and $g^{\alpha\beta}$ that are one the inverse of the other and differential operations $D_{\mu}$ are defined through the connections $\Gamma^{\rho}_{\alpha\beta}$ defined in terms of their transformation law, and because the metric tensors cannot vanish and the connections are not tensors then they will always depend on the particular coordinate system in such a way that there exists a unique local coordinate system in which the metric can be flattened whereas there are in general different local coordinate systems in which one of the different symmetric parts of the connections can be vanished; by demanding the condition of metricity according to the relationship given by $D_{\mu}g=0$ and the Cartan torsion tensor to be completely antisymmetric according to the condition $Q_{\alpha\mu\rho}=Q_{[\alpha\mu\rho]}$ then we have that the local coordinate systems in which the metric tensor is flattened and the symmetric part of the connection vanishes are the same and there is a single local coordinate system in which the only symmetric part of the connection can be vanished: these conditions respectively entail the principle of causality and equivalence, as it has been discussed in \cite{h,a-l,m-l,f/1,f/2}.

We remark that from the metric tensor it is possible to define also the Levi-Civita completely antisymmetric tensor $\varepsilon$; then the completely antisymmetric torsion tensor can equivalently be written in terms of the axial torsion vector as
\begin{eqnarray}
Q^{\beta\mu\rho}=\varepsilon^{\beta\mu\rho\sigma}W_{\sigma}
\label{axialvector}
\end{eqnarray}
and the condition $D_{\mu}\varepsilon=0$ can be inferred as well.

In this background, we will define Riemann curvature tensor $G_{\alpha\beta\mu\nu}$ as
\begin{eqnarray}
G^{\alpha}_{\lambda\mu\nu}=
\partial_{\mu}\Gamma^{\alpha}_{\lambda\nu}-\partial_{\nu}\Gamma^{\alpha}_{\lambda\mu}
+\Gamma^{\alpha}_{\rho\mu}\Gamma^{\rho}_{\lambda\nu}
-\Gamma^{\alpha}_{\rho\nu}\Gamma^{\rho}_{\lambda\mu}
\label{Riemann}
\end{eqnarray}
antisymmetric in both the first and the second couple of indices, allowing only one independent contraction, Ricci curvature tensor $G^{\lambda}_{\alpha\lambda\beta}=G_{\alpha\beta}$, whose contraction is Ricci curvature scalar $G_{\alpha\beta}g^{\alpha\beta}=G$ and this will set our convention.

Riemann curvature tensor, Ricci curvature tensor and scalar, together with Cartan torsion tensor verify
\begin{eqnarray}
D_{\rho}Q^{\rho\mu \nu}+\left(G^{\nu\mu}-\frac{1}{2}g^{\nu\mu}G\right)
-\left(G^{\mu\nu}-\frac{1}{2}g^{\mu\nu}G\right)\equiv0
\label{torsiondiv}
\end{eqnarray}
and
\begin{eqnarray}
D_{\mu}\left(G^{\mu\rho}-\frac{1}{2}g^{\mu\rho}G\right)
-\left(G_{\mu\beta}-\frac{1}{2}g_{\mu\beta}G\right)Q^{\beta\mu\rho}
+\frac{1}{2}G^{\mu\kappa\beta\rho}Q_{\beta\mu\kappa}\equiv0
\label{curvaturediv}
\end{eqnarray}
known as Jacobi-Bianchi identities.

Next we consider that in the case of complex fields also gauge covariance needs to be considered, and in this case differential operations $D_{\mu}$ are defined through the gauge connections $A_{\alpha}$ so to give gauge covariant derivatives.

In this background, we define Maxwell curvature tensor $F_{\mu\nu}$ as
\begin{eqnarray}
F_{\mu\nu}=\partial_{\mu}A_{\nu}-\partial_{\nu}A_{\mu}
\label{Maxwell}
\end{eqnarray}
which is antisymmetric, and thus it is obviously irreducible.

The Maxwell curvature tensor is such that it verifies
\begin{eqnarray}
\partial_{\alpha}F_{\mu\sigma}+\partial_{\sigma}F_{\alpha\mu}+\partial_{\mu}F_{\sigma\alpha}\equiv0
\label{curvaturecgaugerot}
\end{eqnarray}
which are geometric identities known as Jacobi-Cauchy identities, and the commutator of covariant derivatives applied on Maxwell curvature tensor gives
\begin{eqnarray}
D_{\rho}\left(D_{\sigma}F^{\sigma\rho}+\frac{1}{2}F_{\alpha\mu}Q^{\alpha\mu\rho}\right)\equiv0
\label{curvaturegaugediv}
\end{eqnarray}
in the form of conservation laws.

We shall now address the fundamental issue based on the fact that, although in the case of the spacetime curvature (\ref{Riemann}) the object upon which the ordinary derivatives act is a connection and thus it cannot be generalized in order to be written in terms of covariant derivatives, nevertheless in the case of the gauge curvature (\ref{Maxwell}) the object upon which the ordinary derivatives act is a vector and so it could be generalized in order to be written in terms of covariant derivatives as in the following
\begin{eqnarray}
\Phi_{\mu\nu}=D_{\mu}A_{\nu}-D_{\nu}A_{\mu}
\equiv \partial_{\mu}A_{\nu}-\partial_{\nu}A_{\mu}+A_{\rho}Q^{\rho}_{\phantom{\rho}\mu\nu}
=F_{\mu\nu}+A_{\rho}Q^{\rho}_{\phantom{\rho}\mu\nu}
\label{Maxwellgeneralized}
\end{eqnarray}
which is not gauge invariant: however we have also to notice that even if (\ref{Maxwell}) could be generalized up to (\ref{Maxwellgeneralized}), such a generalization is not needed since (\ref{Maxwell}) is already the most general definition if we want Maxwell curvature tensor to be the commutator of gauge covariant derivatives, exactly in the same way in which Riemann curvature tensor is the most general definition if we want it to be the commutator of covariant derivatives, as it has also been discussed in \cite{f/2}.

Within this background, to define matter fields classified according to the value of their spin we have to consider that a given matter field of spin $s$ has in general $2s+1$ degrees of freedom, whose number has to be equal to that of the corresponding $2s+1$ independent solutions of a system of field equations, which has to specify the highest-order time derivative of all the components of the field itself. As it may happen that field equations are not enough to determine the correct rank of the solution, one needs to impose constraints, which are equations in which all components of the field have highest-order time derivatives that never appear; these constraints can be imposed in two ways, either being implied by the field equations or being assigned as subsidiary conditions that come along with the field equations. Although the former procedure seems more elegant, whenever interactions are present it has two types of problems: the first is that the presence of the interacting fields could increase the order derivative of the constraining equation up to the same order derivative of the field equations themselves, creating the possibility that highest-order time derivatives of some component appear to convert the constraint into a field equation, and spoiling the balance between degrees of freedom and independent components of the field; on the other hand if in the constraining equation the highest-order time derivative never appeared or if it actually appeared but could be removed by means of field equations, then the constraint is a constraint indeed, but in this case a second type of problem is that the interacting fields could let terms of the highest-order derivative appear in the equation that determines the propagation of the wave fronts, allowing these terms to influence the propagation of the wave fronts themselves. The equation that determines the propagation of the wave fronts is obtained by considering in the field equations modified by the constraints only the terms of the highest-order derivative, where the derivatives are formally replaced with the vector $n$ getting a matricial equation, of which one has to demand singularity getting what is called characteristic equation; the solutions of the characteristic equation are the normal $n$ to the characteristic surfaces describing the propagation of the wave fronts: if there is no time-like normal then there is no space-like characteristic surface, and therefore there is no acausal propagation for the wave front, as extensively discussed throughout references \cite{v-z/1,v-z/2,p-r} and in the generalized cases in references \cite{d-z,d-w,u-v,n}.

Once this analysis is performed to check causal propagation, the last requirement is that the system of field equations
\begin{eqnarray}
D_{\sigma}F^{\sigma\rho}+\frac{1}{2}F_{\alpha\mu}Q^{\alpha\mu\rho}=J^{\rho}
\label{maxwell}
\end{eqnarray}
with
\begin{eqnarray}
\left(G^{\sigma\rho}-\frac{1}{2}g^{\sigma\rho}G\right)
+8\pi K \left(\frac{1}{4}g^{\sigma\rho}F^{2}-F^{\sigma\mu}F^{\rho}_{\phantom{\rho}\mu}\right)
=-8\pi K T^{\sigma\rho}
\label{einstein}
\end{eqnarray}
and
\begin{eqnarray}
&Q^{\nu\sigma\rho}=16\pi K S^{\nu\sigma\rho}\\
&S^{\nu\sigma\rho}=S^{[\nu\sigma\rho]}
\label{sciama-kibble}
\end{eqnarray}
in terms of the gravitational constant $K$ is postulated so that the torsion and the curvature together with the gauge fields are coupled to the spin $S^{\nu\sigma\rho}$ with the energy $T^{\sigma\rho}$ and also the current $J^{\mu}$, whose conservation laws will be determined by the system of matter field equations, building the set-up of the fundamental field equations at the least-order derivative, as it has also been discussed in \cite{F}.

\subsection{Constraints for the Degrees of Freedom}
Having settled the background in this way, we begin to consider the issue of which matter fields could actually be defined in it; clearly, because the background is characterized by these constraints, then matter fields will behave in a correspondingly restricted way.

\subsubsection{Constrained Matter Field Theories}
In the background thus defined, the complete antisymmetry spin $S^{\nu\sigma\rho}$ with the energy $T^{\sigma\rho}$ and also the current $J^{\mu}$ have to be such that the conservation laws 
\begin{eqnarray}
D_{\rho}S^{\rho\mu\nu}+\frac{1}{2}\left(T^{\mu\nu}-T^{\nu\mu}\right)=0
\label{spin}
\end{eqnarray}
and also
\begin{eqnarray}
D_{\mu}T^{\mu\rho}-T_{\mu\beta}Q^{\beta\mu\rho}-S_{\beta\mu\kappa}G^{\mu\kappa \beta \rho}+J_{\beta}F^{\beta\rho}=0
\label{energy}
\end{eqnarray}
together with
\begin{eqnarray}
D_{\rho}J^{\rho}=0
\label{current}
\end{eqnarray}
are verified when matter field equations are postulated.

We will next consider a few examples of matter fields discussing how they can be defined in order for these constraints to be respected.

\paragraph{Examples of Vector Fields.} In the case of a vector $V_{\mu}$ it is possible to define, beside the standard covariant derivative in terms of the connection, another more special differential operation $Z_{\mu\nu}=\partial_{\mu}V_{\nu}-\partial_{\nu}V_{\mu}$ called exterior derivative with no additional field; although this form of the derivative is compulsory for gauge fields, for a massive vector field there is no gauge symmetry that forces the derivative to be the exterior derivative, and although the exterior derivative is still interesting due to its special form, there is of course the possibility to use the most general covariant derivative defined in the standard way: in the following we will consider massive vector fields whose dynamics is given by covariant derivatives, first in the special case of formal exterior derivative with respect to the most general connection $Z_{\rho\mu}=D_{\rho}V_{\mu}-D_{\mu}V_{\rho}$, and then in the most general case of covariant derivatives $D_{\rho}V_{\mu}$ in the standard form.

Given the complex vector field $V_{\mu}$, the most general Proca matter field equations are
\begin{eqnarray}
D_{\mu}Z^{\mu\alpha}+m^{2}V^{\alpha}=0
\label{generalprocafieldequations}
\end{eqnarray}
in terms of the parameter $m$, which specify the second-order time derivative for only the spatial components but which also develop the constraint
\begin{eqnarray}
m^{2}D_{\mu}V^{\mu}-\frac{1}{2}Q^{\rho\alpha\beta}D_{\rho}Z_{\alpha\beta}
-G^{\alpha\beta}Z_{\alpha\beta}-\frac{i}{2}F^{\alpha\beta}Z_{\alpha\beta}=0
\label{generalprocaconstraints}
\end{eqnarray}
and we see that due to the presence of torsion this constraint contains terms with the second-order time derivative of spatial components, which can anyway be removed by means of field equations, and thus it is a constraint; this constraint allows field equations to specify the second-order time derivative of all components, although by plugging it back into the field equations the presence of torsion gives third-order derivative field equations, which are unacceptable.

Thus a different procedure must be followed, and we can proceed by separating the variables of the vector field according to the usual decomposition that is given by $V_{\mu}=U_{\mu}+D_{\mu}B$ with $D_{\mu}U^{\mu}=0$, employing which the system of the constraint with the system of field equations (\ref{generalprocafieldequations}-\ref{generalprocaconstraints}) is decomposed as
\begin{eqnarray}
\nonumber
&-\frac{1}{2}Q_{\rho\alpha\beta}Q^{\rho\alpha\sigma}D_{\sigma}D^{\beta}B+m^{2}D^{2}B
-\frac{1}{2}D_{\rho}Q^{\rho\alpha\beta}Q_{\sigma\alpha\beta}D^{\sigma}B+\\
\nonumber
&+\frac{1}{2}Q^{\rho\alpha\beta}G_{\sigma\rho\alpha\beta}D^{\sigma}B
-iQ^{\rho\alpha\beta}F_{\rho\alpha}D_{\beta}B
-\frac{1}{2}Q_{\rho\alpha\beta}Q^{\rho\alpha\sigma}D_{\sigma}U^{\beta}-\\
\nonumber
&-D_{\rho}Q^{\rho\alpha\beta}D_{\alpha}U_{\beta}
-iF^{\alpha\beta}D_{\alpha}U_{\beta}
-\frac{i}{2}D_{\rho}Q^{\rho\alpha\beta}F_{\alpha\beta}B
-\frac{i}{2}Q^{\rho\alpha\beta}D_{\rho}F_{\alpha\beta}B+\\
&+\frac{1}{2}F^{\alpha\beta}F_{\alpha\beta}B
+\frac{1}{2}Q^{\rho\alpha\beta}G_{\sigma\rho\alpha\beta}U^{\sigma}
-\frac{i}{2}Q^{\rho\alpha\beta}F_{\rho\alpha}U_{\beta}=0\\
\nonumber
&D^{2}U^{\alpha}-\frac{1}{2}Q^{\mu\sigma\alpha}Q_{\mu\sigma\rho}D^{\rho}B
-D_{\mu}Q^{\mu\sigma\alpha}D_{\sigma}B+iF^{\mu\alpha}D_{\mu}B+\\
\nonumber
&+m^{2}D^{\alpha}B-Q^{\sigma\mu\alpha}D_{\sigma}U_{\mu}
-\frac{i}{2}Q^{\mu\sigma\alpha}F_{\mu\sigma}B+iD_{\mu}F^{\mu\alpha}B-\\
&-G^{\mu\alpha}U_{\mu}+iF^{\alpha\mu}U_{\mu}+m^{2}U^{\alpha}=0
\label{generalprocadecomposedfieldequations}
\end{eqnarray}
in terms of $m$, specifying second-order time derivatives of all components, and without third-order derivatives, and so acceptable.

In this equation, we consider only highest-order derivative, separating the contribution of torsion, the gauge field and the metric, and we write the partial derivatives as formally replaced with $n$ to obtain the matricial equation whose singularity condition
\begin{eqnarray}
n^{2}\left[n^{2}\left(m^{2}+W^{2}\right)-\left|n \cdot W\right|^{2}\right]=0
\label{generalprocacharacteristicequation}
\end{eqnarray}
is the characteristic equation that will have to be discussed.

However such a discussion of the characteristic equation and its solutions is not actually very difficult in this case, because it is always possible to have torsion weak enough to let time-like solutions be present, and so space-like characteristic surfaces can always be achieved; and consequently we have that acausal propagation of the wave front can always be accomplished.

So the matter vector complex field with the exterior derivative has wave fronts which can always be characterized by the acausal propagation.

We will now leave the treatment of the exterior derivatives, to turn our attention to the most general covariant derivative defined in the standard way.

Now given the vector field $V_{\mu}$, the Fermi matter field equations are
\begin{eqnarray}
D^{2}V^{\alpha}+m^{2}V^{\alpha}=0
\label{fermifieldequations}
\end{eqnarray}
in terms of $m$, specifying second-order time derivatives of all components.

We clearly see that because Fermi field equations do not develop constraints, then the characteristic equation has solutions of the light-like type alone.

However precisely because Fermi matter field equations do not develop constraints then subsidiary conditions are imposed to set the degrees of freedom to the $3$ that define vector fields and the set of field equations with constraints is 
\begin{eqnarray}
&D^{2}V^{\alpha}+m^{2}V^{\alpha}=0\\
&D_{\mu}V^{\mu}=0
\label{fermifieldequation}
\end{eqnarray}
where the conserved quantities are given by the current
\begin{eqnarray}
J_{\alpha}=i\left(V^{\mu}D_{\alpha}V^{\ast}_{\mu}-V^{\ast}_{\mu}D_{\alpha}V^{\mu}\right)
\label{fcurrent}
\end{eqnarray}
and by the energy
\begin{eqnarray}
T_{\alpha\nu}=
-g_{\alpha\nu}m^{2}V^{\ast}_{\mu}V^{\mu}+g_{\alpha\nu}D_{\rho}V^{\ast}_{\beta}D^{\rho}V^{\beta}
-D_{\alpha}V^{\ast}_{\theta}D_{\nu}V^{\theta}-D_{\alpha}V^{\theta}D_{\nu}V^{\ast}_{\theta}
\label{fenergy}
\end{eqnarray}
and the spin
\begin{eqnarray}
&S_{\rho\beta\theta}=\frac{1}{2}
\left(V_{\beta}D_{\rho}V^{\ast}_{\theta}-V^{\ast}_{\theta}D_{\rho}V_{\beta}
+V^{\ast}_{\beta}D_{\rho}V_{\theta}-V_{\theta}D_{\rho}V^{\ast}_{\beta}\right)\\
\nonumber
&V_{\beta}D_{\rho}V^{\ast}_{\theta}-V^{\ast}_{\theta}D_{\rho}V_{\beta}
+V^{\ast}_{\beta}D_{\rho}V_{\theta}-V_{\theta}D_{\rho}V^{\ast}_{\beta}+\\
&+V_{\rho}D_{\beta}V^{\ast}_{\theta}-V^{\ast}_{\theta}D_{\beta}V_{\rho}
+V^{\ast}_{\rho}D_{\beta}V_{\theta}-V_{\theta}D_{\beta}V^{\ast}_{\rho}=0
\label{fspin}
\end{eqnarray}
ensuring the complete antisymmetry of the spin, and the spin and the energy with the current are such that conservation laws (\ref{current}), (\ref{energy}) and (\ref{spin}) are verified.
 
For the counting of degrees of freedom, the subsidiary condition in (\ref{fermifieldequation}) already provides the necessary constraints that ensure the right amount of degrees of freedom; but the condition of complete antisymmetry of the spin in (\ref{fspin}) has its only independent contraction given by $V_{\rho}D^{\rho}V^{\ast}_{\theta}+V^{\ast}_{\rho}D^{\rho}V_{\theta}=0$ independent on the subsidiary condition: hence beyond the constraints provided by the subsidiary condition, also $V_{\rho}D^{\rho}V^{\ast}_{\theta}+V^{\ast}_{\rho}D^{\rho}V_{\theta}=0$, not to mention the uncontracted conditions (\ref{fspin}), provide further constraints decreasing the amount of degrees of freedom to less than the $3$ degrees of freedom needed to define the matter vector field as it should. So problems in the number of the degrees of freedom arise.

The matter vector complex field with standard derivative is overconstrained.

So far we have studied the cases of exterior and standard derivatives for the matter vector fields with complex values, where the fact that the matter vector field had complex values was a mere property of the field alone and not of its transformation law, which was still a transformation law of the real Lorentz group; from now on we will turn our attention to the more satisfactory description of matter fields in which the fields are complex fields and for which the transformation law is complex as well, being it a complex representation of the Lorentz group: in the following we will study the case of spinor fields.

\paragraph{Examples of Spinor Fields.} Spinor fields transform according to spinorial transformations; a spinorial transformation $S$ can be expanded in terms of the infinitesimal generators $\sigma^{\mu\nu}=\frac{1}{4}[\gamma^{\mu},\gamma^{\nu}]$ written in terms of the $4$-dimensional matrices $\gamma^{\mu}$ verifying the Clifford algebra and such that from them it is possible to define the matrix $\gamma=i\gamma^{0}\gamma^{1}\gamma^{2}\gamma^{3}$ which is parity-odd. The matrix $\gamma_{0}$ is such that it verifies $\gamma_{0}\gamma_{\mu}^{\dagger}\gamma_{0} =\gamma_{\mu}$ and so used to define $\overline{\psi}=\gamma_{0}\psi^{\dagger}$ the complex conjugate spinor. Spinorial differential operations are introduced as usual.

\subparagraph{Vector-Spinor Fields.} The column of vectors  $\psi^{\mu}$ whose transformation law is given by $\psi'^{\mu}=S\frac{\partial x'^{\mu}}{\partial x^{\nu}}\psi^{\nu}$ defines the Rarita-Schwinger spinors. When these spinors and their derivatives are considered, it is possible to have two types of field equations, according to whether the indices are contracted with the Levi-Civita completely antisymmetric tensor externally or internally between the fields, respectively determining whether the Rarita-Schwinger constraints are implied by the Rarita-Schwinger field equations or must be postulated as subsidiary conditions beside the field equations: in the following we shall consider massive Rarita-Schwinger spinor fields whose dynamics is given by these two field equations, first in the special form that is capable of implying its own constraints, as for the case of the gravitino field, and then in the form in which the constraints must be postulated as subsidiary conditions beside the field equations, as in the original paper by Rarita and Schwinger \cite{d-z,d-w,u-v,n}.

Given the Rarita-Schwinger spinor, the Rarita-Schwinger spinor matter field equations are
\begin{eqnarray}
\varepsilon^{\alpha\nu\rho\sigma}\gamma\gamma_{\nu}D_{\rho}\psi_{\sigma}+
m\sigma^{\alpha\rho}\psi_{\rho}=0
\label{rsfieldequations}
\end{eqnarray}
in terms of the parameter $m$, which specify the time derivative for only the spatial components but which also develop the constraint
\begin{eqnarray}
\nonumber
&4\varepsilon^{\nu\eta\rho\sigma}F_{\eta\rho}\gamma\gamma_{\nu}\psi_{\sigma}
-2i\varepsilon^{\nu\eta\rho\sigma}G_{\alpha\kappa\eta\rho}
\gamma\gamma_{\nu}\sigma^{\alpha\kappa}\psi_{\sigma}
+4i\varepsilon^{\nu\eta\rho\sigma}G_{\alpha\sigma\eta\rho}\gamma\gamma_{\nu}\psi^{\alpha}-\\
&-4i\varepsilon^{\nu\eta\rho\sigma}Q_{\kappa\eta\rho}\gamma\gamma_{\nu}D^{\kappa}\psi_{\sigma}
+3m^{2}\gamma_{\nu}\psi^{\nu}=0
\label{rsconstraint}
\end{eqnarray}
in which because of the presence of torsion there are terms with the time derivative that cannot be removed by using the field equations, so that this constraint is converted into a field equation and the balance between the number of independent field equations and degrees of freedom is lost immediately.

So the matter vector-spinor field in the case in which its field equations imply their own constraints are such that some of their constraints are converted into field equations, spoiling the balance between the number of independent field equations and degrees of freedom right away.

Because this special form of the Rarita-Schwinger equations was given in order to imply its own constraints, which in this case are not gotten anyway, then we do not consider it anymore, and we will turn to consider the Rarita-Schwinger equations with subsidiary conditions.

Given the Rarita-Schwinger spinor, the Rarita-Schwinger spinor matter field equations are
\begin{eqnarray}
i\gamma^{\mu}D_{\mu}\psi^{\rho}-m\sigma^{\rho\alpha}\psi_{\alpha}=0
\label{rsstandardfieldequations}
\end{eqnarray}
in terms of the parameter $m$, which specify the time derivative of all components.

Therefore it is clear that in this form the Rarita-Schwinger equations do not develop any constraint, and henceforth we have that their characteristic equation has solutions that are found to be of the light-like type alone.

Because Rarita-Schwinger matter field equations do not develop constraints, subsidiary conditions must be imposed to set the degrees of freedom to the $4$ that define vector-spinor fields and the set of field equations and constraints is 
\begin{eqnarray}
&i\gamma^{\mu}D_{\mu}\psi^{\rho}-m\sigma^{\rho\alpha}\psi_{\alpha}=0\\
&\gamma_{\mu}\psi^{\mu}=0
\label{rsstandardfieldequation}
\end{eqnarray}
where the conserved quantities are given by the current
\begin{eqnarray}
J_{\nu}=\overline{\psi}^{\alpha}\gamma_{\nu}\psi_{\alpha}
\label{rsstandardcurrent}
\end{eqnarray}
and by the energy
\begin{eqnarray}
T^{\sigma\rho}=
\frac{i}{2}\left(\overline{\psi}^{\alpha}\gamma^{\sigma}D^{\rho}\psi_{\alpha}
-D^{\rho}\overline{\psi}^{\alpha}\gamma^{\sigma}\psi_{\alpha}\right)
\label{rsstandardenergy}
\end{eqnarray}
and the spin
\begin{eqnarray}
&S^{\alpha\nu\mu}=\frac{i}{2}\left[\left(\overline{\psi}^{\nu}\gamma^{\alpha}\psi^{\mu}
-\overline{\psi}^{\mu}\gamma^{\alpha}\psi^{\nu}\right)
+\frac{1}{2}\overline{\psi}^{\sigma}\{\gamma^{\alpha},\sigma^{\nu\mu}\}\psi_{\sigma}\right]\\
&\overline{\psi}^{\nu}\gamma^{\alpha}\psi^{\mu}
-\overline{\psi}^{\mu}\gamma^{\alpha}\psi^{\nu}
+\overline{\psi}^{\nu}\gamma^{\mu}\psi^{\alpha}
-\overline{\psi}^{\alpha}\gamma^{\mu}\psi^{\nu}=0
\label{rsstandardspin}
\end{eqnarray}
ensuring the complete antisymmetry of the spin, and the spin and the energy with the current are such that conservation laws (\ref{current}), (\ref{energy}) and (\ref{spin}) are verified.

Now we will follow another way to count the degrees of freedom, directly comparing the number of constraints: the vector-spinor field is defined with a total number of $4$ constrains upon its components, each being complex, giving a total number of $8$ real constraints; the condition of complete antisymmetry of the spin provides $20$ real constraints, whether the subsidiary condition is considered or not: the degrees of freedom are less than the $4$ degrees of freedom needed to define the matter vector-spinor field. So problems for the Rarita-Schwinger matter vector-spinor field about the right amount of degrees of freedom arise.

The matter vector-spinor field in the case in which its field equations have to be supplemented by constraints as subsidiary conditions is overconstrained.

To this point we have studied cases of matter vector-spinor fields, that is matter fields having both vectorial and spinorial indices; we now turn attention to the case of the scalar-spinor field defined to have a spinorial index alone.

\subparagraph{Scalar-Spinor Fields.} The column of scalars $\psi$ whose transformation law is given by $\psi'=S\psi$ defines the Dirac spinors, as it is widely known \cite{F}.

Given the Dirac spinor, the Dirac spinor matter field equations are
\begin{eqnarray}
i\gamma^{\mu}D_{\mu}\psi-m\psi=0
\label{dfieldequations}
\end{eqnarray}
in terms $m$, with time derivative of all components and unconstrained.

As the Dirac equations are unconstrained it follows that their characteristic equation has solutions of the light-like type alone, again as it is widely known.

On the other hand we have that the entire set of field equations is given by
\begin{eqnarray}
i\gamma^{\mu}D_{\mu}\psi-m\psi=0
\label{dfieldequation}
\end{eqnarray}
and where the conserved quantities are given by the current
\begin{eqnarray}
J_{\nu}=\overline{\psi}\gamma_{\nu}\psi
\label{dcurrent}
\end{eqnarray}
and by the energy
\begin{eqnarray}
T^{\sigma\rho}=
\frac{i}{2}\left(\overline{\psi}\gamma^{\sigma}D^{\rho}\psi
-D^{\rho}\overline{\psi}\gamma^{\sigma}\psi\right)
\label{denergy}
\end{eqnarray}
and the spin
\begin{eqnarray}
S^{\alpha\nu\mu}=\frac{i}{4}\overline{\psi}\{\gamma^{\alpha},\sigma^{\nu\mu}\}\psi
\label{dspin}
\end{eqnarray}
completely antisymmetric, so that this form of the spin with the energy and also with the current is such that conservation laws (\ref{current}), (\ref{energy}) and (\ref{spin}) are verified.

As is it clear the spin is completely antisymmetric automatically, and no further consistency problems arise, so the Dirac spinor is perfectly allowed.

From all these examples, we have shown that no tensorial index can be present, so that we are left with a field having spinorial indices only, that is we have the least-spin field as the only matter field possible within this scheme.

This matter field is the one characterized by causal propagation and the complete antisymmetry of spin achieved in the most general case, while for all conserved quantities the conservation laws are fulfilled as well. 

So we have found that the simplest matter field alone is allowed, all the others less simple matter fields are too complex and thus inconsistent in general, recovering a result alternatively discussed in \cite{F}.
\section*{Conclusion}
In this paper we have extended the Velo-Zwanziger analysis initially focused on the gauge fields also to the metric and especially the torsion field, and we have accompanied it with the study of what possible constraints may arise form the complete antisymmetry of the spin tensor, discussing several examples: we have seen that the matter vector complex field with the exterior derivative has wave fronts with acausal propagation, while the matter vector-spinor field in the case in which its field equations imply their own constraints as for the case of the gravitino field is such that some of these constraints are converted into field equations and some constraints are lost; the matter vector complex field with standard derivative and the matter vector-spinor field that needs to be supplemented by constraints as subsidiary conditions like in the original work of Rarita and Schwinger are both overconstrained; the matter scalar-spinor field has none of these problems regarding the consistency with the constraints.

As a comparison with known previous results, we know the Velo-Zwanziger analysis has shown that gauge curvatures determine propagation with acausal features for the Rarita-Schwinger field of the gravitino and loss of constraints for the spin-$2$ field; in the present treatment, we have shown that if beyond the curvatures there is also torsion then propagation with acausal features is found for the Proca field and loss of constraints for the Rarita-Schwinger field of the gravitino: thus it is clear that there seem to be similar diseases affecting torsionless spin-$\left(s+\frac{1}{2}\right)$ fields and torsional spin-$s$ fields, as if for all these situations torsion would account for an additional spin-$\frac{1}{2}$ component in exacerbating those problems. In addition to the Velo-Zwanziger analysis thus generalized, we have seen that the complete antisymmetry of the spin gives constraints that reduce the number of degrees of freedom; this creates problems in the very definition of the matter fields, as it has already been remarked in \cite{F}. This discussion has shown that the torsion-spin coupling is responsible for either back-reactions that can influence the propagation inducing acausal behaviour or constraints that can reduce the degrees of freedom of the fields spoiling the definition of the fields in exam, giving rise to issues of compatibility with the underlying constraints that are difficult to meet by the complexity of a higher-spin matter field. We have found that the least-spin matter field alone is permitted, while all higher-spin matter fields have problems of consistency in general.

\

\noindent \textbf{Acknowledgments.} This work was financially supported by NSERC Canada.

\end{document}